# Softening online extremes organically and at scale

Elvira Maria Restrepo[1], Martin Moreno[2], Lucia Illari[3], Neil F. Johnson[3,*]
[1]Elliot School, George Washington University, Washington D.C. 20052, U.S.A.
[2]Engineering Department, Universidad de Los Andes, Bogota, Colombia
[3]Physics Department, George Washington University, Washington D.C. 20052, U.S.A.
*neiljohnson@gwu.edu

**Calls are escalating for social media platforms to do more to mitigate extreme online communities whose views can lead to real-world harms[1–37], e.g., mis/disinformation and distrust[38–60] that increased Covid-19 fatalities, and now extend to monkeypox[61], unsafe baby formula alternatives[62,63], cancer, abortions[64,65], and climate change[66-70]; white replacement that inspired the 2022 Buffalo shooter and will likely inspire others[2-15]; anger that threatens elections[71-80], e.g., 2021 U.S. Capitol attack; notions of male supremacy that encourage abuse of women; anti-Semitism, anti-LGBQT hate and QAnon conspiracies[81,82]. But should 'doing more' mean doing more of the same, or something different? If so, what? Here we start by showing why platforms doing more of the same will not solve the problem. Specifically, our analysis of nearly 100 million Facebook users entangled over vaccines and now Covid and beyond, shows that the extreme communities' ecology has a hidden resilience to Facebook's removal interventions; that Facebook's messaging interventions are missing key audience sectors and getting ridiculed; that a key piece of these online extremes' narratives is being mislabeled as incorrect science; and that the threat of censorship is inciting the creation of parallel presences on other platforms with potentially broader audiences. We then demonstrate empirically a new solution that can soften online extremes organically without having to censor or remove communities or their content, or check or correct facts, or promote any preventative messaging, or seek a consensus. This solution can be automated at scale across social media platforms quickly and with minimal cost.**

Online extremes—by which we mean social media communities whose views lie at the extreme of some spectrum, e.g., an anti-vaccination Facebook Page—are the subject of intense attention[1–88]. They now exist across myriad topics and social media platforms[2,3] which suggests that instead of playing continual catch-up with their content as each new topic and crisis emerges, more attention should be paid to the underlying communities themselves. The comprehensive review by Marwick et al.[7] reinforces the fact that most platforms feature in-built community structures such as a Facebook Page, a VKontakte Club, a Gab Group, a Discord Server, a 4Chan Channel, where people can easily aggregate and within which extreme views can hence develop quickly[19,20]. Prior research shows that users within such in-built communities build up a level of trust with their fellow community users, thus tending to pay attention to each others' narratives[89-91]. Moreover, these in-built communities can become interconnected within and across platforms through shared hyperlinks, meaning they can quickly share, escalate, and generate new extreme content and associated dis/misinformation for any new crisis. Already regarding Ukraine, Facebook has failed[92] to label 80% of the articles that promote the narrative that the U.S. is funding use of bioweapons.

Online extreme communities can generate physical harms in addition to broader distrust, anger and hate, especially when their ideas spillover to the millions in mainstream communities to whom they may be connected[1-37]. Federal Drug Administration commissioner Dr. Robert Califf has suggested that their health mis(dis)information is now the leading cause of death in U.S. while President Biden stated[93] that their activity on Facebook is "killing people". It reduced Covid-19 vaccine uptake in the mainstream population and a growing rejection of mRNA technology now threatens new vaccines for future pandemics[44]. While only 22 people saw the Buffalo shooter's video live, an interconnected online network of extreme communities has enabled millions of mainstream users to see it since[9]. The 2022 Antisemitism Worldwide Report showed anti-Semitic incidents escalated globally in the last year and that this was fueled by a network of radical left- and right-wing extremes on social media[81,82]: social media played "an exceptionally alarming role" with numbers in New York and Los Angeles alone, almost doubling from the previous year. Political extremists carried out approximately 450 murders in the U.S. in the past decade and have fueled mass-scale violent protests, e.g., Jan 6 U.S. Capitol attack[72] and increasing use of Facebook by separatists to incite widespread violence and ethnic hatred in Nigeria[74]. Following the Buffalo massacre, President Biden called for an end to the "relentless exploitation of the internet to recruit and mobilize terrorism". The Buffalo shooter, like others before, cited extreme online communities as the origin of his radicalization, not people he met personally, leading New York Governor Kathy Hochul and others to directly blame social media platforms[94].

Online extremes persist despite ongoing platform intervention efforts, and despite an explosion in the number of related research papers, conferences, journalistic reports, daily website compilations by NGOs, and even academic journals, involving researchers from across disciplines: from communications and media studies through law, sociology, political science, and engineering to computer science and physics[69]. Added to this are new initiatives from scientific bodies to attack the problem using core science, e.g., American Physical Society[95], the AAAS[96], the U.S. National Academies of Science[97], plus dedicated studies from the American Academy of Political and Social Science[97]. The sheer number of these works makes it impossible to provide a concise review that is fair to all researchers and disciplines. Hence, we defer a 185 page (yet still incomplete) literature overview to the Supplementary Information (SI), and instead discuss some solutions that have been proposed.

One proposed approach to mitigating online extreme communities is to demand more transparency about the social media algorithms that can facilitate their growth. Twitter's likely new owner Elon Musk is moving in that direction and there are new, open decentralized social media structures such as Manyverse and Bluesky. But having open-source software could introduce new security risks while doing little to boost overall transparency[99]. Also, more transparency of social media data might threaten Americans protections under the Fourth Amendment or laws like the Stored Communications Act. In short, transparency from most platforms is unlikely to be forthcoming.

Another proposed approach is through government legislation and rules on social media platforms, with severe financial and even criminal penalties. On 23 April 2022, the EU announced new legislation which holds platforms financially liable for extreme content that is deemed harmful. This Digital Services Act will impose significant moderation requirements on social media companies. The U.S. lags behind on this, in part due to the broader implications of having such laws and the existence of the First Amendment right to free speech. The Platform Transparency and Accountability Act was introduced in December 2022 followed by the May 2022 United States Senate Committee on the Judiciary Subcommittee on Privacy, Technology, and the Law. However even if this progresses, the lack of detailed information about underlying algorithms means a purely legislative approach runs the risk of creating new laws and regulations that do not adequately address harms, or inadvertently making problems worse[1].

Yet another approach is to tackle the news sources that feed them. There are many new technology tools such as NewsGuard, a prominent web browser extension that embeds source-level indicators of news reliability into users' search engine results, social feeds, and visited URLs. However, Aslett et al. have shown that credibility labels on news have very limited effects on the quality of news diet and do not counteract misperceptions[100].

Given these uncertainties, it appears the only chance of a quick but scalable solution is for social media platforms themselves to increase their efforts. But given their ultimately finite resources, should platforms be planning to do more of the same? Or is there something new that they can do that has empirical support and is scalable?

The answer to the first question as suggested by our analysis of the Facebook ecology in Figs. 1 and 2, is that doing more—even much more—of the same will not work. Facebook, the largest social media platform, announced a significant increase in the amount of its interventions starting November 2020 with a focus around vaccines and Covid-19[16]. As outsiders, we cannot know exactly what Facebook did and when, however we can assess the net impact of its interventions by analyzing the temporal evolution of its online ecology. Specifically, we take the pre-Covid ecology published in Ref. 56 (Fig. 2(a)) showing the battle over vaccine views between interconnected communities (Pages) and we update it through 2022. We then look for significant changes up until the November 2020 ramp-up (Fig. 2(b)), and from then to now (Figs. 2(c)(d)).

We refer to Methods for the methodology[56], and our later discussion of limitations. The SI contains data for the nodes (each node is a Facebook Page) and links (node A links to node B when Page A follows Page B and hence content flows from community B to community A). Since 2019, the content in this ecology of nearly 100 million users expanded beyond vaccines to include Covid-19 treatments and policies (extracts available upon request). We use the ForceAtlas2 network layout[56] in which nodes repel each other while links act as springs: hence nodes that are visually closer to each other have environments which are more highly interconnected—and hence are more likely sharing and discussing the same material—while nodes that are far apart are not. The layout is color agnostic, i.e., the segregation of colors (node categories) is an entirely spontaneously effect and is not pre-determined by the algorithm.

Figure 1's upper left cartoon shows the two main categories of Facebook interventions that we detect since November 2019: (1) removing anti-vaccination community accounts (red nodes) and their content, particularly from November 2020 onwards (Fig. 2(c)) and (2) posting factual messaging into these communities (blue ring) to try to counteract their extreme views. This milder intervention (i.e. (2)) appears to have been favored during early Covid since the number of nodes did not change significantly between Fig. 2(a) and 2(b) but the fraction of nodes containing the Facebook intervention message did (Fig. 2(b)).

Figure 1 (left panel) lists 4 substantial problems that our analysis detects and that still persist, with some being a direct consequence of these Facebook interventions. Hence Facebook doing more—even much more—of the same in the future will not work.

Problem l is that even though Facebook has removed some large and active extreme communities (nodes), operationally the ecology remains largely unchanged because of a hidden self-repair effect. Figure 2(c) shows nodes removed following the November 2020 ramp-up, and those that felt sufficient moderator pressure to self-remove (i.e., become private). Since these post ramp-up removals (Fig. 2(c)) represent a cross-section of the full ecology of extreme and neutral nodes shown in Fig. 2(b), one would expect this to have had a major impact on the network's ability to share content. However, Fig. 2(d) shows there was negligible impact on the subnetwork of top 20 nodes ranked by highest betweenness centrality, which is a measure of a node's (community's) ability to act as an efficient conduit of content. Red nodes (anti-vaccination communities) are the most prominent and highest ranked within these top 20: while many were removed by Facebook (Fig. 2(d) upper panel), there was a system rewiring such that other red nodes (anti-vaccination communities) took their place (Fig. 2(d) lower panel). A similar resilience effect to Fig. 2(d) is seen from November 2019 to November 2020 (not shown). Putting this together with the fact that a larger than expected number of links appear between the red nodes shown in Fig. 2(d) despite there being no rule that high betweenness centrality nodes are necessarily highly interlinked between themselves, means that there is a self-repairing core 'mesh' of extreme anti-vaccination communities within the full ecology (Fig. 2(b)) that have—and can continue to—share and distribute extreme content not only with each other but also with the other nodes in the full ecology, including the millions of users in mainstream neutral communities with whom they are connected (Fig. 2(b)), e.g. parenting communities.

Problem 2 is that a major Facebook messaging intervention continues to miss key audience sectors and is generating ridiculed. The message (red ring, Fig. 1 upper left) was posted at the top of many communities' Page, directing its users to the official health authority, e.g., CDC in U.S., NHS in U.K. The white ring in Fig. 2(b) encircles approximately 90% of the ecology's nodes (communities) that received this posted message. While it covers the core region of red nodes, it misses many outside that are directly connected to mainstream neutral communities, as well as missing all the mainstream neutral communities themselves, e.g., parenting communities. The innovative use of clothing size to ridicule this messaging (Fig. 1 upper left) likely caught the attention of many women and mothers across both mainstream neutral and extreme communities.

Problem 3 is that a key piece of these online extremes' narratives has, and is still, being mislabeled as incorrect science. The ubiquitous story of microchip sensors (quantum dots) being embedded in vaccines is completely false—but not because the science is impossible, but rather because there is absolutely no intent to do so. A major journal published[101,102] a proof-of-principle of the science in December 2019 (title shown) just before Covid-19 vaccine discussions started. This led to media headlines (title shown) that got picked up by extreme communities, together with the fact that funding came from the Gates Foundation and China. When media, fact-checkers and moderators, label this as impossible science, the extreme communities point their users (and mainstream users) to these scientific publications to support their claims of an establishment cover-up.

Problem 4 is that the increased threat from Facebook's intervention ramp-up post-November 2020, is inciting extreme communities—and now mainstream parent-centric ones as alluded to by the graphic—to create parallel presences on other platforms. Not only does this give their supporters the impression they are under attack because they are 'on to something', but it can end up exposing them to a broader range of extreme audiences, e.g., 'save the children' is a popular narrative for QAnon communities on Gab.

Figure 1's middle panel highlights why a more hands-off intervention approach as adopted during early Covid-19 until November 2020, also will not work. This period of lighter intervention crudely approximates the scenario that some have suggested of letting natural evolution take its course. The positions and sizes of the guide-to-the-eye rings in Figs. 2(a) and (b) are identical in order to visually illustrate the observed tightening: specifically, there is internal strengthening of the subnetwork of extreme communities as well as strengthening of connections between anti-vaccination communities and different categories of neutrals (non-red and non-blue), e.g., parenting communities. Following the ball-and-spring layout, the observed tightening indicates increased likelihood of sharing the same (extreme) content. Hence returning to pre-November 2020 levels of intervention or hoping that a natural evolution of views can take its course, will not work.

Our proposed solution (Fig. 3) is a scalable scheme that softens online extremes via organic deliberation in anonymous, heterogeneous groups formed online around a given topic. The Methods gives details of our proof-of-principle experiments, while Extended Data Figure 1 shows some early implementations in which we co-located participants to troubleshoot technical issues (they still interacted in their groups anonymously online). Each set of experiments involved nearly 100 volunteers. They start by answering an online question on some topic (e.g., vaccines, government policy) to determine their view on a spectrum. Simple software then assigns them into heterogeneous groups. They answer this same question when they exit their group. All answers and identities are kept secret. Each group is assigned an online space in which its members interact anonymously (e.g. 5 per group, see Methods). They are all given the same simple prompt on the topic of interest to stimulate interactions. There is no moderation and no set duration. Activity in each group tends to die out within an hour. As in everyday online reality, conversations can wander and involve sarcasm as well as deliberation; some people may dominate, and some may drop out.

An average softening of extremes is observed across groups as in Fig. 3(b)(d), and across cohorts. In general, the more heterogeneous the groups, the more softening, though many more experiments are needed to establish any rigorous quantitative relationship. We checked that this softening is not an aberration of the online technology, by manually repeating the experiment offline (Fig. 3(f)).

There are several reasons this experiment can be operationalized quickly at scale across social media platforms, to provide a perpetual automated softening scheme with low overhead cost:
(1) Even though as-yet unknown external factors may reduce the softening at any given time for any given topic or platform or groups, the automated process of forming anonymous heterogeneous groups can operate continually and in parallel within and across platforms: hence even small, intermittent softenings will tend to add up over time and at scale. (2) While our experiment required answering an initial question to construct heterogeneous groups, online users have already crudely done this step by choosing a particular community type, e.g., red, blue etc. in Fig. 2(b). Knowing from our ecology maps the links between communities of different types (e.g., red linked with blue in Fig. 2(b)) and hence who is likely listening in on (or at least familiar with) who, means that platforms can inject a simple automated invitation into them both to draw in participants and hence form groups. Participant uptake does not have to be high since each community can contain thousands of users and yet each group needs only a few (see Methods). Prior research[89-91] suggests that any subsequent modified views will likely be listened to within their respective communities, while experiments of Centola[103] et al. suggest an entire community could eventually flip its type with approximately 25% committed views. Even if the ecology map is not well-known, the typical breadth of the distribution of views (Fig. 3(b)(d)) means that fairly heterogeneous groups can still be produced even if the selection is effectively random and hence the scheme can still operate.

Computer predictions from applying our scheme to the current vaccine view ecology (Fig. 3(g)) are shown in Extended Data Figure 2. The simulations assume the scheme is applied in a repeated automated way using softening values per group similar to Figs. 3(b)(d). The typical duration of one hour per group engagement is used as the timescale, and groups operate in parallel. The simulations predict substantial softening of extremes at scale within a few weeks.

The theoretical foundation for our solution stems from the deliberation work of Steiner[104–107] and others[108–111] who focused on face-to-face deliberation, not at scale online (see Methods). It is also consistent with mathematical results[5] showing how extreme communities evolve in the face of a 'limited (American) pie', a motivating concept reinforced recently by Kayyem to explain support for white replacement[112]. It also connects to 'nudge influence' and 'nudgeability' analyses of de Ridder et al. and others[91] as well as shifts in group beliefs[39], though this solution is a form of spontaneous collective nudging with no notion of nudging or prompting based on a ground truth or accuracy.

Limitations of this paper include the fact that our analysis of interventions (Fig. 1, 2) focuses on Facebook and Pages. However, most social media platforms have similar in-built community structures (e.g., a channel/group on Telegram or Gab, a club on VKontakte, a Page on Facebook), hence we expect our findings to apply more broadly. Moreover, links to private spaces such as Facebook Groups and private chat or streaming apps, often appear within these public communities (see Extended Data Figure 3) suggesting that the visible Pages ecosystem can act as a crude proxy for such private networks. The classification terms 'anti' and 'pro' are limited and relative, e.g., pro-X could in reality be anti-'anti-Y'. Indeed, "mixed, unstable or unclear" ideologies now account for more than half of all anti-radicalization referrals in the U.K. and "mixed ideologies is where it's all heading"[113]. However even if a multi-dimensional classification were adopted, these communities can still be seen as extreme and hence occupy non-binary locations on the surface of some circle or sphere etc., versus the two ends of a line. Limitations of our solution (Fig. 3) include the fact that we cannot guarantee polarization in a specific group would not increase because of some unknown external factor. However, the key to this solution lies in the accumulation of many small softenings over time from operating repeatedly at scale across platform(s).

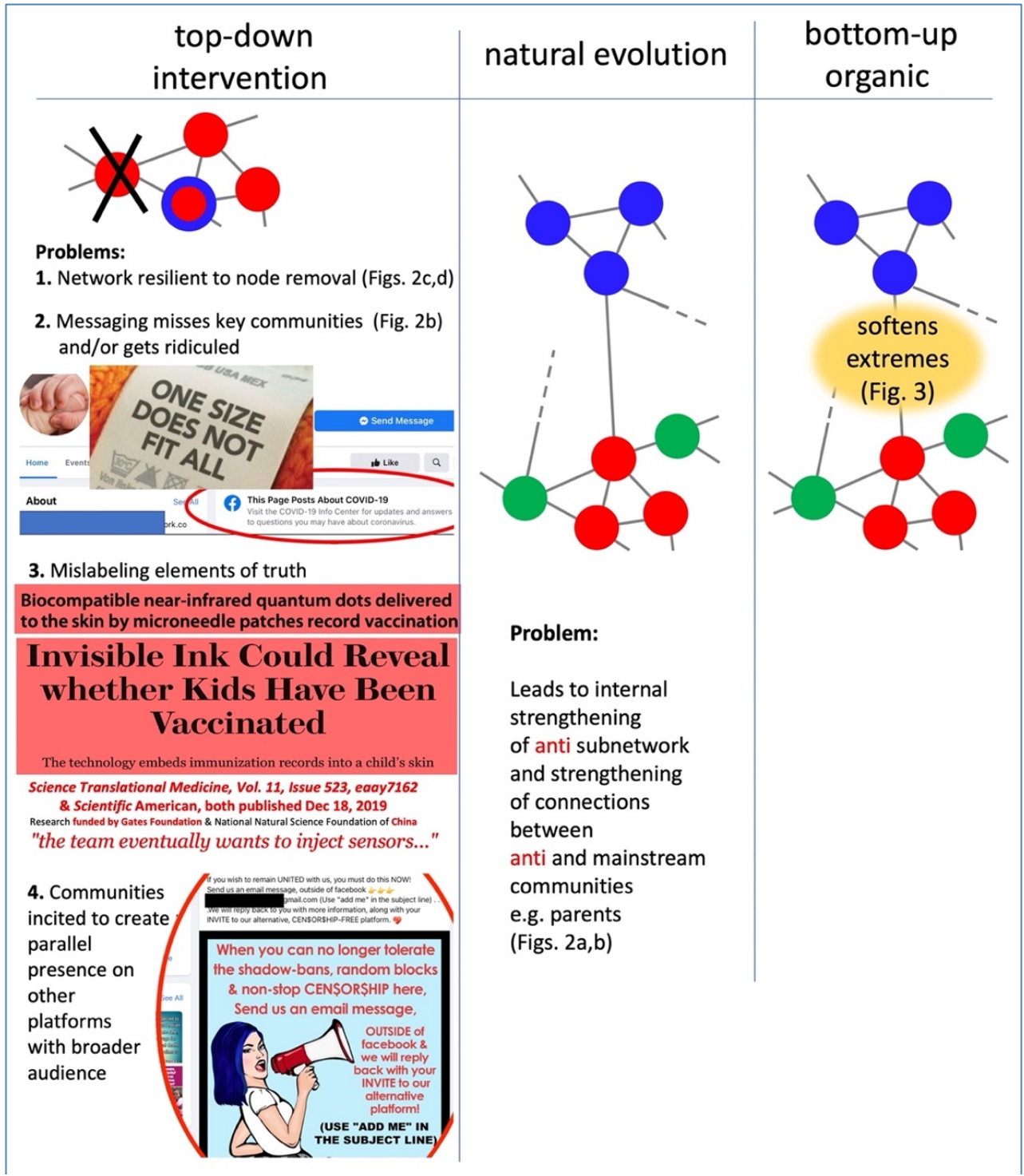

**Fig. 1: Problems with interventions to date, and the new organic scheme (Fig. 3). Examples in left and middle panels come from the Facebook ecology surrounding vaccines[56] that we updated through 2022. As in Fig. 2, each red node is a Page (i.e., in-built community) classified as anti-vaccination, each blue node is a Page classified as pro-vaccination, other (e.g., green) nodes are Pages classified as neutral. Each in-built community (i.e., Page, node) can contain 10-1,000,000+ like-minded supporters. The headlines shown in Problem 3 are actual titles taken from the named scientific publications.**

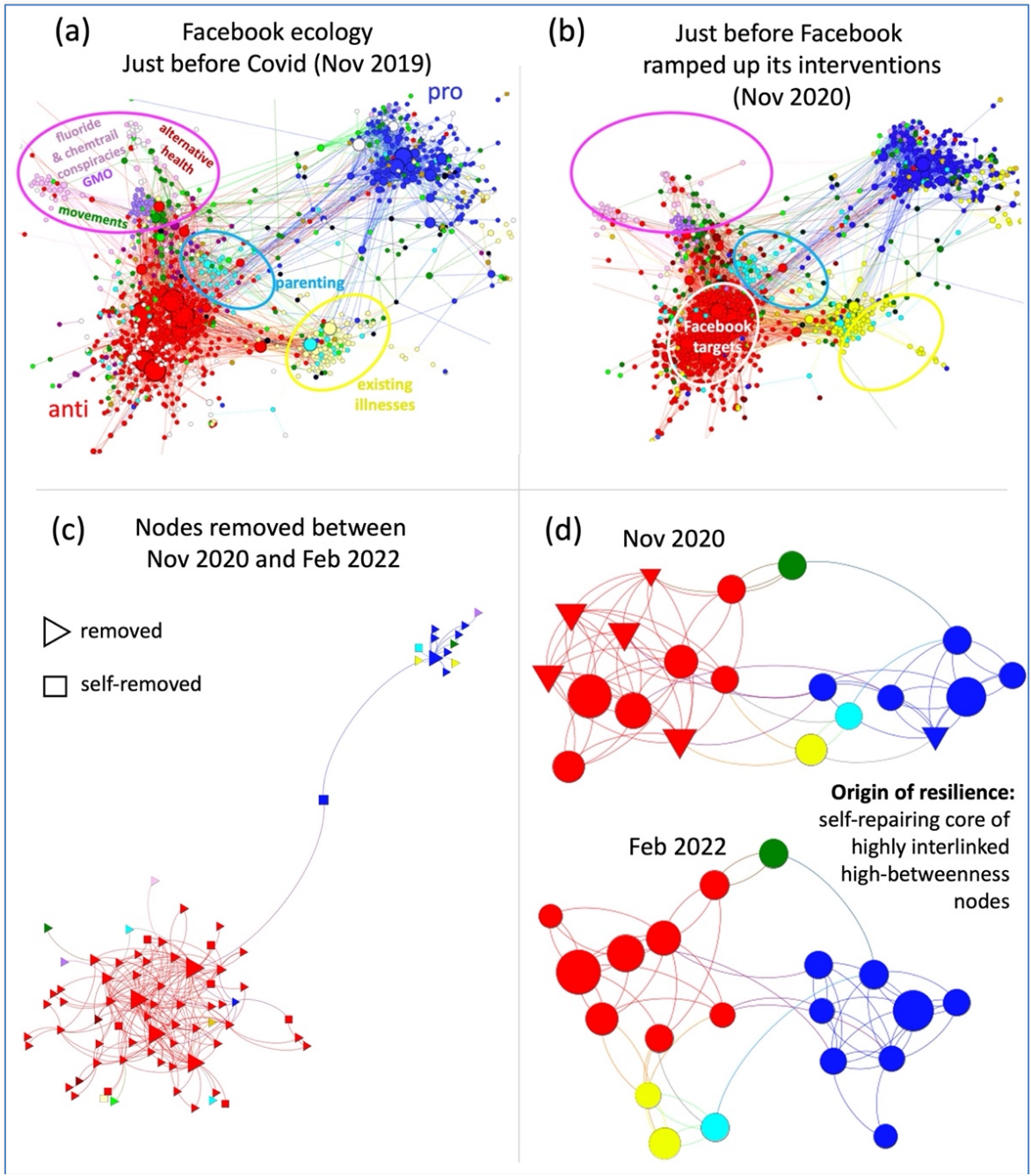

**Fig. 2:** Changes over time in the 2019 Facebook vaccine-view ecology[56] shown in (a), containing nearly 100 million users. Node size represents betweenness values, i.e. a node's ability to act as a conduit of content. Starting November 2020, Facebook ramped up its interventions, notably removals (see (c)). Until then (i.e. (a) and (b)) it focused on posting factual messages (Fig. 1 left panel, blue ring and example shown). ForceAtlas2 layout mechanism means nodes (i.e., Pages) appearing closer together are more likely to share content. (a) and (b) have same scale: rings indicating categories of neutral Pages are in same position to highlight strengthening of connections within reds and between reds and neutrals. Though shown here by different colors to distinguish categories, all neutral nodes are green nodes in Fig 1. (d) shows self-repairing core 'mesh' of high betweenness anti-vaccination (red) communities from within (b) that have—and can continue to—share and distribute extreme content with each other and with other nodes in full ecology, including millions of mainstream users in neutral nodes. The specific nodes in this core mesh (Fig. 2(d)) change over time, but its structure remains robust.

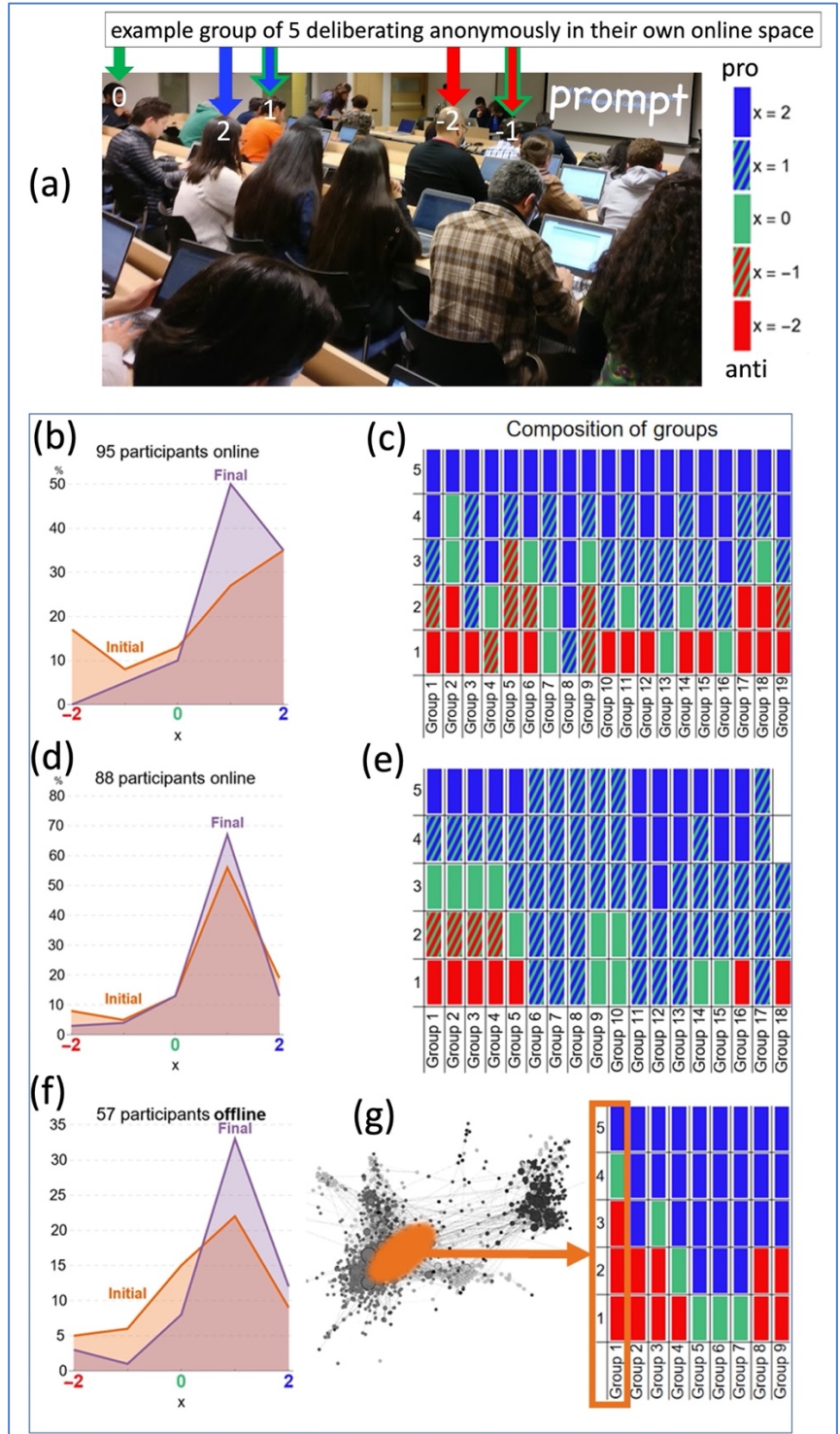

**Fig. 3: Experimental evidence for our scalable solution that softens online extremes via organic deliberation in anonymous, heterogeneous groups formed online around a given topic (see Methods). Small group size (e.g. 5) encourages participation and happens to mimic modern group chat apps. (a) shows a photo from an early cohort in which participants, though interacting anonymously online, were co-located to troubleshoot technical problems. (b)-(e) shows representative outcomes from different cohorts. (f) shows an equivalent offline version we ran to check softening is not a technological artefact. (g) shows how the scheme can be implemented, guided by the ecology maps from which heterogeneous groups can be formed (see Extended Data Figure 2).**

**Methods**

**1. Data analysis and network construction (Figs. 1 and 2).** See SI for data and replication files. We start with the Facebook Page data from the authors of Ref. 56 which showed the vaccine view ecology as of November 2019. We then updated their data and network through 2022 using their same methodology[56]. Each node in the network ecology is a Facebook Page (i.e., an in-built Facebook community) and is either anti-vaccination (red node), pro-vaccination (blue node) or some other neutral category based on their content (e.g., parenting). A link from Page A to Page B means Page A explicitly lists Page B as one of the Pages to which it links, not necessarily because it agrees with Page B's content but because Page B's content is of interest to Page A. Such a link could have appeared because Page A's users noticed it and then recommended it to Page A's managers who then established the link. Though Facebook has changed the appearance of its interface, these Page-level links still persist in the background. Such a link from Page A to Page B creates an information conduit feeding content from Page B to Page A, exposing A's users to B's content. The detailed methodology providing the data from Ref. 56 starts with a core of nodes (Facebook Pages discussing vaccines) obtained manually. Each node (Page) and all its content can be seen publicly and none of the analysis requires individual-level information. The content of the nodes is then labeled manually, with analysts working independently and afterwards checking for consensus. Without requiring any discussion, consensus was reached at this first stage in approximately 80% of the cases. Subsequent cases were then discussed until agreement was reached. An expanded list of nodes is then built by looking to see which nodes each node linked to (i.e. Pages to which each Page linked itself). This established the first step in a systematic, iterative process for building a master set of nodes and links. Links between nodes are directed. While other definitions of nodes and links are possible, this one avoids the need for individual-level information and makes the definition of each community (node) unique since each Page has its own unique identification number. The authors of Ref. 56 pruned the link list manually in each iteration since they were aiming at extracting meaningful links rather than the default of a nearly fully connected network with potential links on all possible topics. Because this process after several iterations yields of order $\approx$1000 nodes (communities), but each contains of order 100,000+ users, the resulting network is visually manageable and interpretable, as opposed to being overwhelmed with links or being too sparse. This means that the open-source software Gephi and its ForceAtlas2 algorithm provides an uncluttered spatial representation, as shown in Figs. 2(a)(b). It also means that the network is scalable to the population level, since 1000 nodes with 100,000+ users each means potentially observing the dynamics and behavior of 1000 x 100,000+ = 100+ million individuals. The betweenness centrality of a node is a standard network measure that quantifies the ability of that node to act as a conduit of information: this quantity is discussed in detail, with simple examples, in network science textbooks.

2. **Experiments (Fig. 3).** Extended Data Figure 1 shows illustrative pictures from some of our earlier experiments in which we had co-located volunteers in case of any technical problems. We carried out multiple experiments: each involved a cohort of nearly 100 volunteers. Many took place in Colombia because of the ease of setting up, the ease of obtaining approval from the institutions, and the explicit interest and support of these institutions as well as from the incumbent President of Colombia (evidence available from authors). As further quality reassurance, an entire experiment was overseen in real time by an experienced academic who is now a member of Facebook's Oversight Board. Most of the early locations from which volunteers were obtained, were universities where the volunteers were students, faculty, and administrative staff. Other sets of experiments drew in high school students. Some involved business-people in commercial settings. Details of locations etc. and lists of experiments including extracts of anonymous chats in each group, are available from the authors upon reasonable request. In all cases, the volunteers were not paid. Examples of the cohort of approximately 100 volunteers, include (1) students, faculty and administrative staff at the Universidad de Los Andes in Bogota, Colombia. These covered a wide range of ages and of economic backgrounds; (2) high school students at a school in a rural area outside the capital city; (3) business-people in a major global city. Softening on average across groups was observed in each case, as illustrated in Fig. 3. The volunteers did not need any particular educational background, and they covered a wide range of ages from teenagers to seniors. The idea behind our experiments was first floated in Ref. 111 but without the current platform implementation under conditions of online deliberation in small heterogeneous groups. The simple software that we wrote for the experiments in this paper to create online spaces and assign participants from across the distribution of views to these groups, is not special in any way. Any webchat page could be used and any few-line code for selecting groups. The necessary process to create such webchat pages can be easily automated by any social media platform at negligible cost.

We chose small groups to encourage individuals to interact more and hence avoid the outcome of having many people ending up sitting rather passively in some larger group discussion. It also mimics the small size of groups in popular group chat apps. Specifically, our debriefings with previous participants suggested that having roughly 5 participants is a good compromise number per group, hence we chose 5 (Fig. 3). We additionally ran an offline version of the experiment. The similarity of the offline results (Fig. 3(f)) to the online ones (Fig. 3(b)(d)) suggests that the softening observed in the online scheme is not due to some spurious technical quirk.

Our experiments also satisfy all the minimum requirements (Steiner Refs. 104-107, or basic norms of Habermas Refs. 108-110) that have been shown as key to effective deliberation. These are: voluntary participation; reciprocity; active participation of most within each group (i.e. at least 3 in each group or else the group was shut down); and a robust question/issue that produced different positions/opinions. We tried different classification questions and prompts to identify topics that would engage the volunteers across the pro/neutral/anti spectrum. The initial question that we used to score views on the spectrum (Fig. 3(a)) was one that seemed to engage people the most (i.e., had less volunteers immediately leave): it concerned support (or not) of the controversial peace agreement that was proposed following the end of Colombia's multi-decade guerilla war. While the topic of peace had been politicized prior to this proposed agreement, the complexities of the agreement itself stimulated wide public debate. Importantly for our study, this controversial topic of the peace agreement is just like a topic such as vaccines (Fig. 2) in that we observe in the online content that is produced by the groups, a mix of rational arguments and also emotions including distrust, anger and fear -- exactly as observed in the content produced by the vaccine-view ecology in Fig. 2. So although the topics of peace and vaccines are of course very different, the collective dynamic of views is similarly charged. The fact that the distributions in Fig. 3 are not perfectly symmetric, suggests that our cohorts had some implicit bias toward 'pro' (i.e., support the proposed peace agreement). However, this does not affect the conclusions of our study, i.e., there is softening of the extremes.